\documentclass[]{emulateapj}

\usepackage{times,graphicx,url}
\usepackage{amssymb,amsmath}
\usepackage{aas_macros}
\usepackage{natbib}
\usepackage{color}
\usepackage{tabularx}

\newcommand{\wfirst}{\emph{WFIRST}}

\newcommand{\mabuls}{\emph{MaB$\mu$LS}}

\newcommand{\thetae}{\theta_{\mathrm{E}}}
\newcommand{\re}{r_{\mathrm{E}}}

\newcommand{\tein}{t_{\mathrm{E}}}

\newcommand{\tzero}{t_{\mathrm{0}}}

\newcommand{\uzero}{u_{\mathrm{0}}}

\newcommand{\mearth}{M_{\earth}}

\newcommand{\xzero}{x_{\mathrm{0}}}
\newcommand{\yzero}{y_{\mathrm{0}}}
\newcommand{\uc}{u_{\mathrm{c}}}
\newcommand{\ucmax}{u_{\mathrm{c,max}}}
\newcommand{\tc}{t_{\mathrm{c}}}
\newcommand{\xc}{x_{\mathrm{c}}}
\newcommand{\yc}{y_{\mathrm{c}}}
\newcommand{\croin}{CROIN}
\newcommand{\close}{^{\mathrm{c}}}
\newcommand{\res}{^{\mathrm{r}}}
\newcommand{\wide}{^{\mathrm{w}}}
\newcommand{\umax}{u_{0,\mathrm{max}}}

\begin{document}

\title{Speeding up low-mass planetary microlensing simulations and modeling: the Caustic Region Of INfluence}
\shorttitle{Caustic Region of Influence}
\shortauthors{Penny}

\author{Matthew T. Penny}
\affil{Department of Astronomy, Ohio State University, 140 West 18th Avenue, Columbus, OH 43210, USA}
\email{penny@astronomy.ohio-state.edu}
\date{\today}

\begin{abstract}
Extensive simulations of planetary microlensing are necessary both before and after a survey is conducted: before to design and optimize the survey and after to understand its detection efficiency. The major bottleneck in such computations is the computation of lightcurves. However, for low-mass planets most of these computations are wasteful, as most lightcurves do not contain detectable planetary signatures. In this paper I develop a parameterization of the binary microlens that is conducive to avoiding lightcurve computations. I empirically find analytic expressions describing the limits of the parameter space that contain the vast majority of low-mass planet detections. Through a large scale simulation I measure the (in)completeness of the parameterization and the speed-up it is possible to achieve. For Earth-mass planets in a wide range of orbits it is possible to speed up simulations by a factor of ${\sim} 30$--$125$ (depending on the survey's annual duty-cycle) at the cost of missing ${\sim} 1$~percent of detections (which is actually a smaller loss than for the arbitrary parameter limits typically applied in microlensing simulations). The benefits of the parameterization probably outweigh the costs for planets below $100\mearth$. For planets at the sensitivity limit of AFTA-WFIRST, simulation speed-ups of a factor ${\sim} 1000$ or more are possible.
\end{abstract}

\section{Introduction}\label{intro}

Gravitational microlensing searches for planets are beginning to yield statistically interesting sample sizes~\citep{Gould2010,Cassan2012} that are set to increase significantly with the advent of new and proposed surveys on the ground~(MOA-II--\citealt{MOAIIref}, OGLE-IV--\citealt{OGLEIVref}, KMTNet--\citealt{KMTref}) and in space (Euclid--\citealt{Penny2013}, WFIRST--\citealt{Spergel2013}). Full understanding of the results of these surveys and the planning of optimal observing strategies requires extensive, computationally-expensive simulations for the calculation of detection efficiencies or yield predictions.

The major bottleneck in the simulation of planetary microlensing is the computation of lightcurves. To compute the magnification of the binary microlensing event with a finite source for a single data point requires either the solution of multiple complex fifth-order polynomials, or costly inverse ray shooting. Each lightcurve typically consists of thousands of data points, and each simulation typically requires thousands to millions of trial lightcurves to be generated in order to obtain reasonable Poisson uncertainties due to the low per-event probability of planet detection.

Significant effort has been invested in increasing the speed at which one can compute the base unit of the lightcurve -- the binary lens magnification. Various approaches have been taken to compute finite source magnifications, either by contour integration combined with numerical solution of the lens equation~\citep{Gould1997, Dominik1998, Bozza2010}, by inverse ray shooting~\citep{Rattenbury2002, Dong2006, Bennett2010a} or by a hybrid of the two~\citep{Dong2006, Dominik2007a}. Others have dug deeper and improved the efficiency of the basic numerical functions these are built upon~\citep{Skowron2012}. Further gains are made by avoiding finite source calculations where approximations will suffice, e.g., the hexadecapole approximation~\citep{Pejcha2009, Gould2008}.

Perhaps more important in reducing the computational expense of microlensing calculations is the choice of an efficient parameterization. The majority of work on this front has focused on the problem of fitting an observed microlensing event with a binary lens model. In this situation, most of the parameter space is uninteresting -- one would prefer to avoid a brute force exploration of parameter space and go straight to the desired answer. This is typically achieved by using a parameterization that matches the features of the event being studied in order to reduce correlations between parameters and decrease the parameter-space volume that must be searched \citep[e.g.][]{Albrow1999, An2002, Cassan2008, Bennett2012}. Simulations to determine detection efficiencies or predict yields have the opposite goal of exploring the entire parameter space. The specialized parameterizations developed for modeling observed events are often difficult to apply to such brute force parameter explorations because the prior distributions of specialized parameters (e.g., caustic crossing durations) are not easily mapped from the more fundamental parameters \citep[e.g., Einstein crossing timescale, mass ratio and projected separation][]{Cassan2010}.

Despite the desire to search the entire parameter space, it is known that most of it contains uninteresting microlensing events that do not show signs of the planets that orbit the lens. If it is possible to identify this region of parameter space before calculating the lightcurve, it is possible to avoid costly lightcurve calculations and significantly speed-up planetary microlensing simulations. This paper introduces a new parameterization (the Caustic Region Of INfluence, or CROIN) that makes it possible to easily identify uninteresting regions of parameter space, whilst keeping a set of geometric parameters with uniform prior distributions. This allows remarkable efficiency gains in the simulation of low-mass planetary microlensing, up to factors of a thousand in reduced computation time. The structure of the paper is as follows. Section~\ref{method} describes the standard binary microlensing parameterization, the new CROIN parameterization and how to transformation between them. In Section~\ref{speedup} we estimate the speed-up that the \croin{} parametrization enables, before assessing its accuracy and any biases it introduces in Section~\ref{accbias}. In Section~\ref{discussion} we discuss first the region of parameter space in which the \croin{} parametrization is useful, before discussing its possible applications. We conclude in Section~\ref{conclusion}.

\section{The method}\label{method}

Before describing the \croin{} parameterization, we first review the standard parameterization and the three caustic topologies that are possible for a binary lens. We conclude the section by describing how to convert between the two parameterizations.

\subsection{The standard parameterization}

The lightcurve of a microlensing event is determined by the magnification pattern of the lens and the trajectory of the source passing through it. The magnification pattern of a binary lens is determined by just two parameters: $s$, the projected separation of the lens components in units of the Einstein ring radius $\re$, and $q=M/M_*$, the mass ratio of the components, where $M$ is the mass of the planet and $M_*$ is the mass of the primary. The source trajectory is described by three parameters that have uniform distributions: $\tzero$, the time of closest approach of the source to a reference point; $\uzero$, the impact parameter at closest approach (normalized to the Einstein radius); and $\alpha$, the angle made by the source trajectory relative to binary axis, which points from the primary to the planet. Finally, the Einstein crossing timescale $\tein$, is the time taken for the source to travel one angular Einstein radius $\thetae$. In this work we choose the reference point as the position of the primary lens, though there are a proliferation of preferred reference points, each with their own advantages and disadvantages.

\subsubsection{The range of $\uzero$ and $\tzero$}

When simulating microlensing events one is forced to choose a range for the parameters $\uzero$ and $\tzero$ from which to draw uniformly, which is usually a compromise between efficiency and completeness. The choice for $\tzero$ is relatively straightforward, and is usually the range of time over which data will be taken, possibly with some margins outside this range to capture planetary signals in the wings of events just outside the observing window. The choice for the maximum value of $|\uzero|$, $\umax$ is more complicated. The probability of detecting a planet falls as $\uzero$ increases, but is finite for $\uzero\lesssim |s-1/s|$ \citep[where $s-1/s$ is the approximate position of the planetary caustics,][]{Han2006}, which given projection effects can become large means that one is forced to compromise between completeness and computation time. A choice of $\umax=3$ is common for planetary microlensing simulations, but inevitably some fraction of planet detections are lost (in Section~\ref{accbias} we find ${\sim} 5$~percent for $\umax=3$ and ${\sim} 20$~percent for $\umax=1$).

\subsection{The Caustic Region of Influence (CROIN) parameterization}\label{definecroin}

The deviations of the binary lens magnification pattern from that of a single lens are strongest at the caustic curves~\citep{Schneider1986} and fall off rapidly outside the caustic~\citep[e.g.,][]{Gaudi2002a, Gaudi2002b}. The caustics are therefore a natural choice for the reference point of an improved impact parameter-based parameterization.

In the CROIN parameterization we define the time of closest approach $\tc$ and impact parameter $\uc$ with respect to a reference point $(\xc,\yc)$ centred on the planetary caustic(s). With the number, position and size of caustics differing for each lens topology, we must therefore choose a different reference point and impact parameter range $\ucmax$ for each topology. We treat each topology separately below. Because we will use the center of the planet signature as our reference point, we assume the range of $\tc$ will be the range of time over which data will be taken; this assumption relies on the planetary signature being short compared to the season length, which we show is justified in Section~\ref{accbias}.

\subsubsection{Defining  $\ucmax$}

\begin{figure}
\includegraphics[width=\columnwidth]{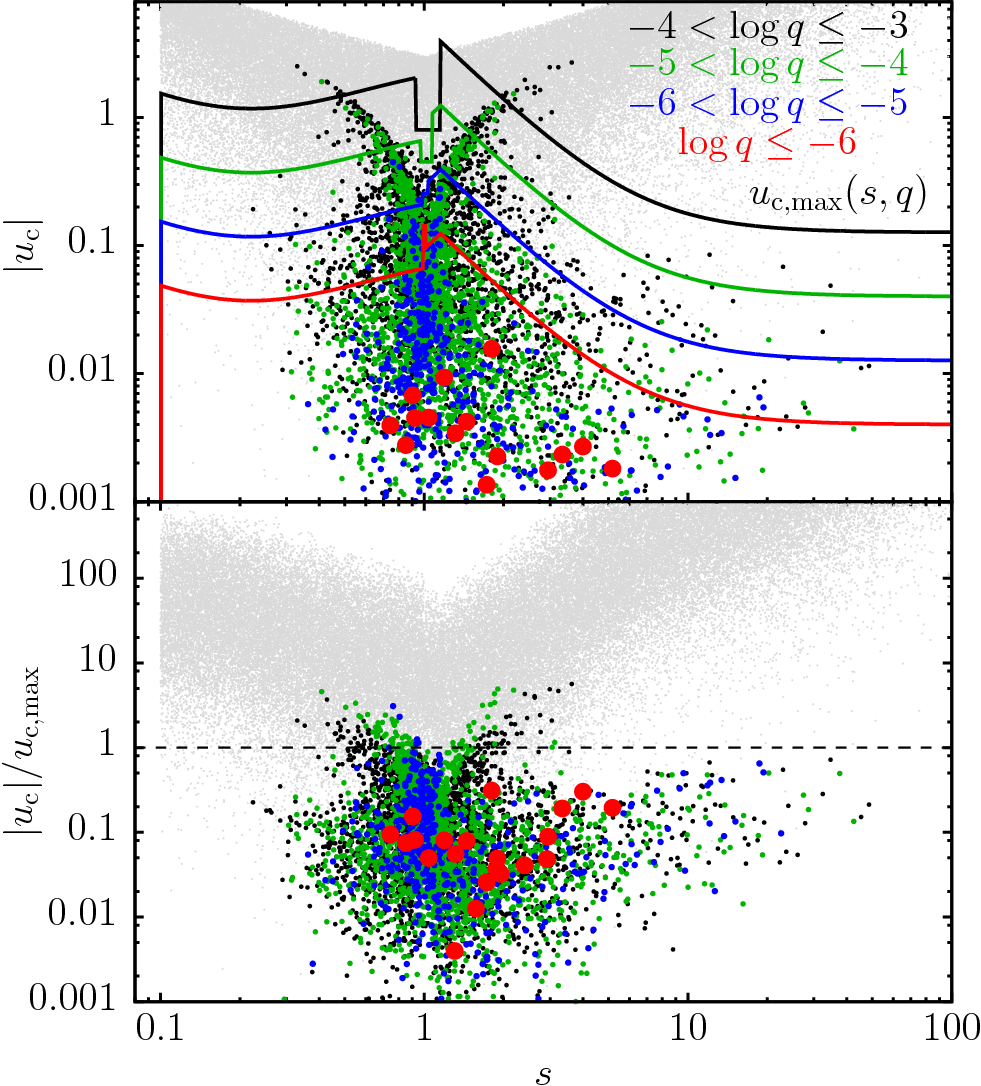}
\caption{\emph{Top panel}: The CROIN impact parameter $\uc$ plotted against projected separation $s$ for a large number of simulated planetary microlensing events. Gray points are a sample (${\sim}1$\%) of all simulated events, while colored points are events where the planet was detected, with colors coded to different ranges of mass ratio $q$ (because of the smaller number of detections with $\log q\le -6$, we use a larger point size to emphasize them). Note that except for the ``goat-horn'' features, most detections lie below the grey non-detections. The colored lines show our adopted limits for the size of the CROIN $\ucmax$, with the line drawn at the maximum of the $q$-range corresponding to its color, i.e., the black line is $\ucmax(s,q)=10^{-3}$. \emph{Bottom panel}: The same data, but now for each point $\uc$ is normalized by $\ucmax$.}

\label{croindef}
\end{figure}

In order to define expressions for $\ucmax$ we employed a plot of $\uc$ against the projected separation $s$, as shown in Figure~\ref{croindef}. For this we used a simulation of WFIRST-AFTA from \citep{Spergel2013} covering a large, but non-uniform range of planet mass ratio.\footnote{Specifically, the planet semimajor axis was drawn from a log-uniform distribution and the planet mass was drawn from the \citet{Cassan2012} mass function that is forced to saturated at $5.5\mearth$ ($2$~dex$^{-2}$~planets per star) as described by \citet{Penny2013}. This mass function, coincidentally, roughly cancels out the microlensing detection efficiency (until it saturates) and so is useful for producing a roughly uniform number of planet detections as a function of mass. This choice of mass function here has no impact on the results.} Once the reference point is defined (see the following subsections), the standard $(\uzero,\tzero)$ parameters are easily transformed to $(\uc,\tc)$ as described in Section~\ref{conversion}. The plot is shown in the top panel of Figure~\ref{croindef}, with colored points marking planet detections of differing mass ratio, and light grey points marking a small fraction of the simulation's non-detections. Here and throughout the paper we consider a planet to be detected if it causes a $\Delta\chi^2>160$ between the fit of a single point lens lightcurve model and the true lightcurve, which is standard for simulations of space-based microlensing surveys (see \citet{Yee2013} for a discussion).

\begin{figure}
\includegraphics[width=\columnwidth]{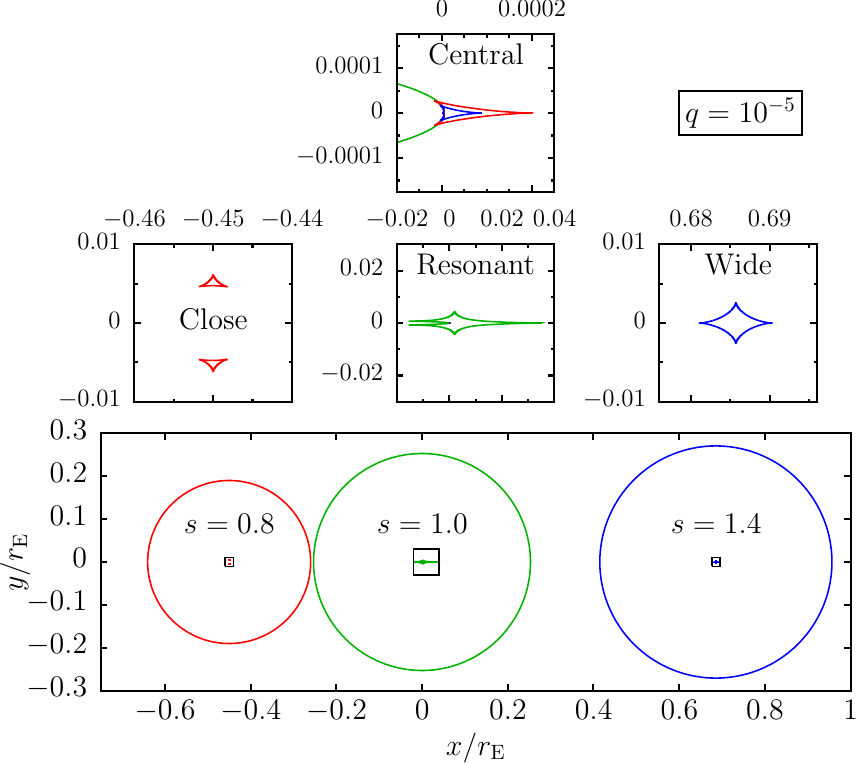}
\caption{Plots showing the caustics for lenses with close (red), resonant (green) and wide (blue) topologies for a lens with mass ratio $q=10^{-5}$. In each case the primary lens is located at $(x,y)=(0,0)$ and the planet is located at $(+s,0)$. The middle panels zoom in on the boxes shown in the lower panel. The central caustics of the close and wide topology lenses are plotted, but are too small to be resolved, in the central panel, so the top panel zooms in on the central panel by more than a factor of 170 to reveal their structure. The circles in the lower panel show the size of the \croin{} for each example caustic topology.}
\label{caustics}
\end{figure}

The goal of plotting $\uc$ against $s$ is to identify a functional form for $\ucmax$ as a function of $s$ and $q$ above which there are very few detections. This is very similar to the process one follows when using the rejection method for drawing random deviates from a general probability distribution function~\citep{Press1986}, except that here we do not need to guarantee that the ``comparison'' function is always greater than the probability density function. The first attempt at defining a functional form used analytic approximations of the size of the caustics scaled by a constant factor to include planet detections via non-caustic-crossing trajectories, but we found that these did not fit the upper envelope of detections very well, regardless of the scaling factor. The final form for $\ucmax$ was found by modifying these expressions with further analytic functions of $s$ and $q$ for each topology. Several iterations through candidate functional forms and their numerical parameters were made before arriving at those presented below. The following three subsections present the final forms for the reference points and $\ucmax$ for each caustic topology in turn and justify the choices made. Figure~\ref{caustics} shows examples of the CROIN for a planet with $q=10^{-5}$ relative to the caustics for each topology. 

\subsubsection{Close topology}

The lens has a close topology when the condition
\begin{equation}
\frac{q}{(1+q)^2} < s^{-8}\left(\frac{1-s^4}{3}\right)^3
\end{equation}
is satisfied~\citep{Erdl1993}. The close topology lens has 3 caustics: a central caustic near the primary and two identical planetary caustics off the binary axis. The most prominent feature of close lenses with low mass ratios is an elongated region of demagnification on the binary axis between the two planetary caustics. The positions of the planetary caustics can be very accurately approximated analytically~\citep{Bozza2000a, Han2006}, therefore we choose the reference point for the close topology to be
\begin{equation}
(\xc\close,\yc\close) = \left(\frac{1}{1+q}\left[s-\frac{1-q}{s}\right],0\right),
\label{cref}
\end{equation}
where we have used Bozza's slightly more accurate expression for the $x$ position of the caustics and the superscript c represents the close topology. Similarly, the superscripts w and r will represent wide and resonant topologies respectively. 

The size of the \croin{} that we choose for the close topology is
\begin{equation}
\ucmax^{\mathrm{c}} = \left\{\begin{array}{lr} 0 & \text{if}\: s<0.1\\ \left(4 + 90s^2\right)\frac{\sqrt{q}}{s\sqrt{1+s^2}} & \text{otherwise}\end{array}\right..
\label{ucmaxc}
\end{equation}
For very small separations, $s<0.1$ we have assumed that there will be no planet detections, which is reasonable even for massive planets. For larger separations, the term in square brackets is a modifier and the fractional term is an analytic estimate for the separation between the two planetary caustics~\citep{Han2006}. We chose the separation of the planetary caustics to set the \croin{} size because the demagnification region lies between these two caustics. The constant term in the modifier accounts for the fact that the demagnification region is elongated along the binary axis somewhat (i.e., actually larger than the separation between the caustics). The strong $s^2$ term attempts to grow the \croin{} to include the central caustic as the projected separation approaches resonance and the central caustic grows and begins to cause a significant number of detections in relatively high-magnification events. Detections via the central caustic form the ``goat horn'' feature in the scatter plot of detections in Figure~\ref{croindef}, and it can be seen that for planets with $q\gtrsim 10^{3.5}$ the $s^2$ growth term will begin to significantly degrade the efficiency gains of the \croin{} parametrization, but not so for lower mass ratios where the $\sqrt{q}$ term keeps $\ucmax<1$.

\subsubsection{Wide topology}

The lens has a wide topology when~\citep{Erdl1993}
\begin{equation}
s > \sqrt{\frac{(1+q^{\frac{1}{3}})^3}{1+q}}
\end{equation}
The wide topology has two caustics: a central caustic like the close topology lens and a single planetary caustic that lies on the binary axis. We choose the center of the planetary caustic as our reference point. Again accurate analytical expressions for this position exist \citep{Bozza2000, Han2006}
\begin{equation}
(\xc\wide,\yc\wide) = \left(s-\frac{1}{[1+q]s},0\right).
\end{equation}

We choose the \croin{} size for wide topologies to be
\begin{equation}
\ucmax^{\mathrm{w}} = \left[4 + \min(90s^2,160s^{-2})\right] \sqrt{q}.
\label{ucmaxw}
\end{equation}
The $\sqrt{q}$ term is simply the size of the Einstein ring of the planet at large $s>>1$. Similar to the close topology, contours of the deviation from the magnification of a point lens are elongated along the binary axis, and a similar constant plus growth term modifier is applied. For $s>\smash{\left(\frac{4}{3}\right)}^2$ the $160 s^{-2}$ term is used to grow the CROIN as $s$ shrinks closer to resonance, but below this the \croin{} is large enough to cover the central caustic and the $90 s^2$ term begins to shrink the \croin{} as $s$ decreases further in a way that approximately matches the slope of the upper envelope of the right-hand goat horn.

\subsubsection{Resonant topology}

The resonant topology only has a single caustic near the primary lens. While the planetward (positive $x$) side of the resonant caustic is larger in extent than the side closest to the primary that extends to negative $x$, it is significantly weaker, and negative perturbations to the magnification pattern extend away from the caustic on the negative side. We therefore choose the primary lens position as the reference point
\begin{equation}
(\xc\res,\yc\res)=(0,0).
\label{rref}
\end{equation}

\begin{figure}
\includegraphics[width=\columnwidth]{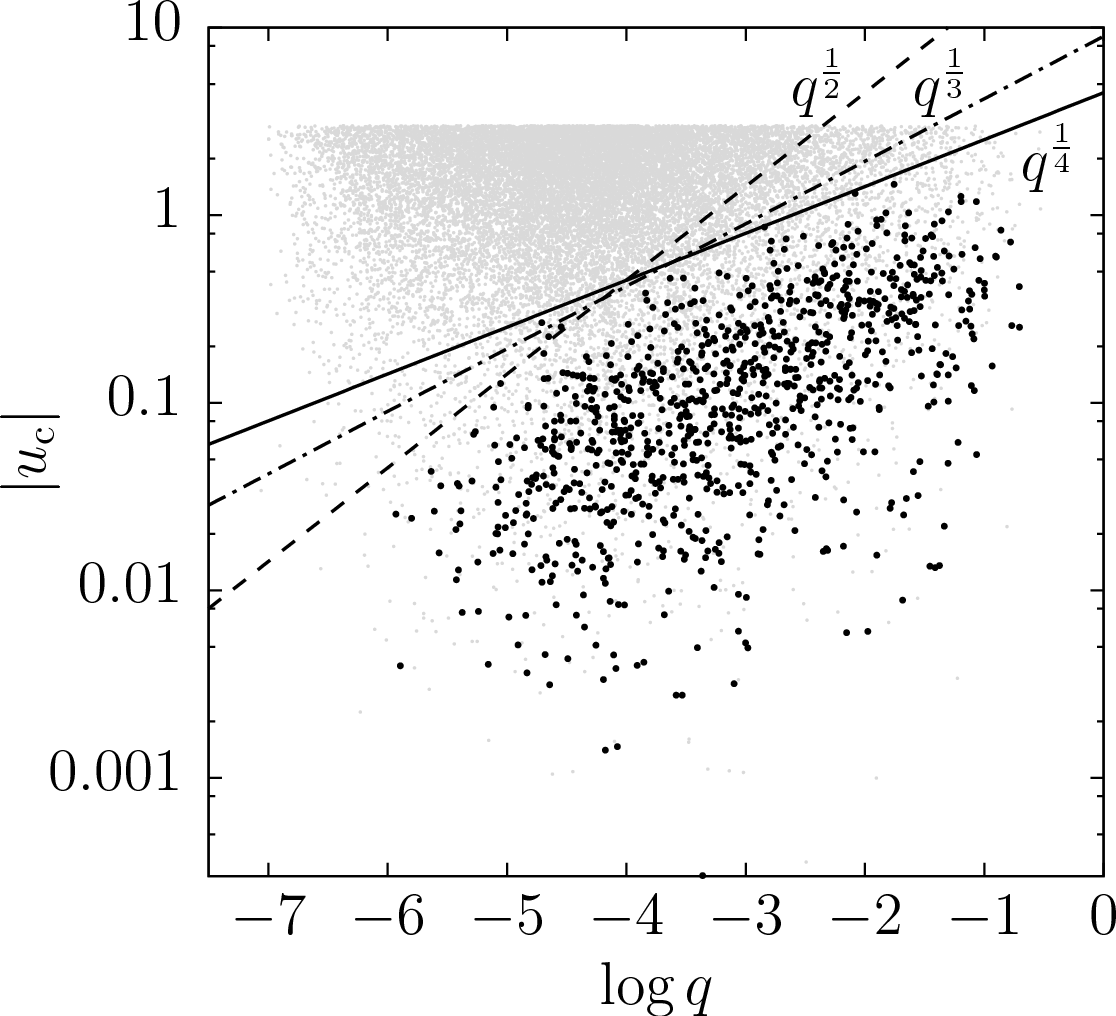}
\caption{Plot of the \croin{} impact parameter $\uc$ against mass ratio $q$ for resonant topology lenses from Figure~\ref{croindef}. Gray points are non-detections while black points are detections. Lines show potential power laws that could be used to describe the upper envelope of detections. The solid line, $|\uc|=4.5q^{1/4}$, shows the actual choice for $\ucmax\res$.}

\label{resonant}
\end{figure}

The \croin{} size for resonant lenses is chosen to be
\begin{equation}
\ucmax\res = 4.5q^{\frac{1}{4}}.
\label{ucmaxr}
\end{equation}
There is only a small range of $s$ with a resonant configuration for small mass ratios, so there is little use in attempting to find a dependence on $s$. We found the simple scaling by examining the detections in Figure~\ref{resonant} and fitting a power law by eye to their upper envelope. A $q^{1/3}$ power law fits the distribution as well as a $q^{1/4}$, but we choose the latter in order to be lenient at low $q$ where our statistics are poor.

\subsection{Converting between parameterizations}\label{conversion}

Both the \croin{} and standard parameterizations are impact parameter-based. Converting the parameters of one to that of the other is simply a case of moving the origin. We provide the more general conversion for a \croin{} center that may lie off the binary axis at $(\xc,\yc)$, though in all the cases we consider here the center of the CROIN lies on the binary axis. The off-axis situation may occur if you wished to extend the \croin{} parameterization to smaller separations by taking the center of the secondary caustics as the \croin{} center, or were cosnsidering multiplanet systems where it is no longer possible for all the planets to lie on the same axis.

\begin{figure}
\vspace{12pt}
\includegraphics[width=\columnwidth]{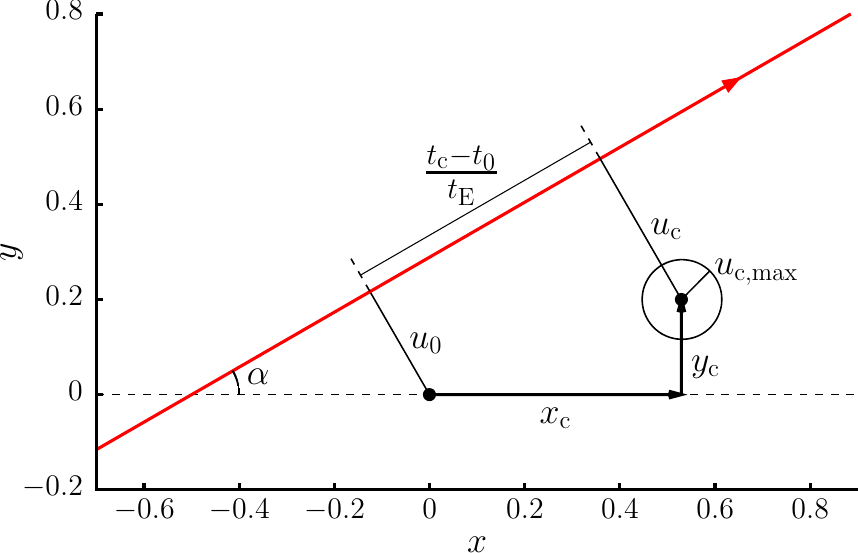}
\caption{Schematic diagram of the $(\uzero,\tzero)\rightarrow (\uc,\tc)$ coordinate transformation. The red line shows the source trajectory, while black lines define the geometry. For the sake of completeness we show the general case of the caustic that lies off the reference axis. We adopt the standard convention that the impact parameter is positive if the source passes the reference point in a clockwise fashion.}
\label{coordtrans}
\end{figure}

Figure~\ref{coordtrans} shows the geometry of the coordinate transformation. From this geometry it is relatively straight forward to derive the following expressions
\begin{equation}
\frac{\tc-\tzero}{\tein} = \xc\cos\alpha + \yc\sin\alpha,
\end{equation}
\begin{equation}
\uc-\uzero = -\xc\sin\alpha + \yc\cos\alpha.
\end{equation}
These expressions are general and apply to the transformation between any two impact parameter-based parameterizations. If one's prefered reference point for the standard parameterization differs from the position of the primary lens, the above formulae can be adapted with the further transformations $\xc\rightarrow(\xc-\xzero)$ and $\yc\rightarrow(\yc-\yzero)$, where $(\xzero,\yzero)$ is the position of the alternate reference point relative to the primary lens.

\section{Speed-up using CROIN}\label{speedup}

\begin{table}
\caption{Speed-up as a function of $s$ and $q$}
\label{qsspeedup}
\begin{tabularx}{\columnwidth}{@{}Xcccccccc}
\hline
\hline
& \multicolumn{8}{c}{$\log s$}\\
$\log q$ & -0.75 & -0.50 & -0.25 & 0.00 & 0.25 & 0.50 & 0.75 & 1.00 \\ 
\hline
-2 & {1.8} & {1.2} & {\bf 1.0} & {\bf 2.1} & {\bf 1.0} & {2.0} & {8.9} & {26} \\
-3 & {6.7} & {2.9} & {\bf 1.9} & {\bf 3.8} & {\bf 1.7} & {5.3} & {28} & {83} \\
-4 & {20} & {8.0} & {\bf 6.0} & {\bf 6.6} & {\bf 5.5} & {15} & {88} & {265} \\
-5 & {67} & {24} & {\bf 18} & {\bf 11} & {\bf 17} & {48} & {276} & {854} \\
-6 & {208} & {76} & {\bf 57} & {\bf 20} & {\bf 55} & {155} & {922} & {2733} \\
-7 & {691} & {248} & {\bf 193} & {\bf 36} & {\bf 179} & {462} & {2785} & {8628} \\
\hline
\end{tabularx}
{\bf Note:} Assumes 100~percent duty cycle and $\umax=3$; speed-up increases for lower duty cycles. Bold values indicate the traditional lensing zone~\citep{Wambsganss1997}.
\end{table}

The simplest way to define and estimate the speed-up that the \croin{} allows over the standard parameterization is to ask what fraction of events drawn from the standard parameter range fall within the \croin{} parameter range. The speed-up is the inverse of this fraction. This can be computed trivially by Monte Carlo sampling without actually simulating events. We did this for a range of $s$ and $q$ and show the results in Table~\ref{qsspeedup}. In these computations we drew $\uzero$ uniformly from the range $-\umax$ to $\umax$ with $\umax=3$, $\alpha$ uniformly from $0$--$2\pi$ and assume that there is a $100$~percent duty cycle, i.e., that the microlensing events peak within the period when the survey is observing (the observing season). This may not be a good assumption for surveys with short seasons, as we discuss in Section~\ref{biases}. In the traditional lensing zone~\citep[$0.6\lesssim s\lesssim1.6$,][]{Wambsganss1997} the speed-up is fairly modest until $q\le 10^{-5}$. At large separations, $s\gtrsim5$, the speed-up is always considerable, and for low mass ratios it is extreme.

\section{Accuracy and Bias}\label{accbias}

Our goal now is to assess the accuracy of the \croin{} parameterization, and make ourselves aware of any biases in the parameter distributions of detected planets that it introduces. To assess bias we just want to measure accuracy as a function of various parameters.

The accuracy and bias is impossible to measure without reference to some observational setup, because this is what determines whether a planet will be detected or not. The simulations described in this paper are of the WFIRST-AFTA mission, which is likely to be the most sensitive microlensing survey it is cost-effective to perform, so the accuracy of the \croin{} parameterization for other surveys is likely to be higher than that which we determine for WFIRST-AFTA. While one may consider the absolute number of detections per unit time to be the best measure of survey sensitivity, for the purposes of assessing the accuracy of the \croin{} parameterization this is not actually the case, because the total number of detections is primarily determined by the number of microlensing events that are monitored. Instead, it is the noise floor and the cadence that determine the most subtle features that can be detected. Of the currently imagined microlensing surveys, WFIRST-AFTA undoubtedly has the lowest noise floor thanks to its space-based photometry and high-resolution (minimizing the effect of blending). In the simulations here we have assumed a noise floor of $1$~mmag, which would be extemely difficult to achieve from the ground. In terms of planned cadence, WFIRST at $15$~min has one of the highest cadences for a microlensing \emph{survey}\footnote{Current ground-based follow-up networks achieve cadences ${\sim 1}$~minute when observing extremely high-magnification events~\citep[e.g., ][]{Dong2009}, but the \croin{} parameterization was not designed for high-magnification events, which are well served by the standard parameterization anyway.}, though KMTNet has a slightly higher cadence at $10$~minutes. This is not a sufficiently large difference to counter WFIRST's space-based advantage, though we will discuss how to adjust the \croin{} for different cadences in the next subsection. 

The simulation data we used in Section~\ref{definecroin} does not have enough low-mass planet detections to properly assess any biases that using the \croin{} introduces, so we ran significantly larger simulations. These simulations were too computationally expensive to cover a large range of masses, so we only performed simulations of planets at two fixed masses: $1$ Earth mass and $100$ Earth masses. For each planet mass we performed two simulations. The first drew event parameters from the standard parameterization, and we used this to compute the accuracy of and biases introduced by the \croin{} parameterization; we refer to this simulation as the \emph{standard} simulation. The second simulation (referred to as the \emph{CROIN} simulation) drew event parameters using the \croin{} parameterization, and we used this to compute the accuracy of the standard parameterization -- i.e., to estimate the number of planet detections that are missed by imposing a cut on $\uzero$.

\begin{table}
\caption{Simulation parameters}
\label{simparam}
\begin{tabularx}{\columnwidth}{lXX}
\hline
\hline
Paramter & & \\
\hline
Survey & \multicolumn{2}{c}{WFIRST-AFTA} \\
Cadence & \multicolumn{2}{c}{15~min}\\
Season length & \multicolumn{2}{c}{72~d}\\
Seasons & \multicolumn{2}{c}{6}\\
Out-of-season gaps & \multicolumn{2}{c}{${\sim}110$~d or ${\sim}840$~d}\\
Noise floor & \multicolumn{2}{c}{1~mmag}\vspace{6pt}\\
& Standard & \croin{} \\
$\uzero$ range & $-3 \rightarrow 3$ & No limit \\
$\tzero$ range & $0 \rightarrow 2010$ & No limit \\
$\uc$ range & No limit & $-\ucmax \rightarrow \ucmax$ \\
$\tc$ range & No limit & In season \\
\hline
\end{tabularx}
\end{table}

These simulations are essentially identical to the \wfirst{} simulations in \citep{Spergel2013}, and which will be presented in detail in Penny et al. (in prep.). Full details of the simulation mechanics are given by \citep{Penny2013}. The important details of the simulations are summarized in Table~\ref{simparam}.

\begin{table}
\caption{Incompleteness of each parameterization}
\label{croinacc}
\begin{tabularx}{\columnwidth}{@{}Xcc}
\hline
\hline
& \multicolumn{2}{c}{Incompletness $f_{\mathrm{missed}}$} \\
Mass & \croin{} & Standard\\
\hline
$1\mearth$ & $1.34\pm0.22$~\% &  $6.88\pm0.10$~\% \\
$100\mearth$ & $6.87\pm0.91$~\% & $5.22\pm0.33$~\% \\
\hline
\end{tabularx}
\end{table}

We measure the accuracy of each parameterization relative to the other by defining the incompleteness $f_{\mathrm{missed}}$. For the \croin{} parameterization we measure it with respect to the standard parameterization, i.e., $f_{\mathrm{missed}}$ is the fraction of planet detections drawn from the standard parameterization limits for which the \croin{} parameters are within the \croin{} parameterization limits. We do the opposite to measure the standard parameterization incompleteness by running a simulation and drawing $\uc$ and $\tc$ uniformly and weighting events appropriately to take into account the variable $\ucmax$. Our results are summarized in Table~\ref{croinacc}. We find that for $1$-$\mearth$ planets, the \croin{} parameterization has $1.3$~percent incompleteness relative to the standard parameterization. For $100$-$\mearth$ planets the \croin{} parameterization's incompleteness is $6.9$~percent. The \croin{} completeness relative to the standard parameterization is worth comparing with its opposite, the standard completeness relative to the \croin{} parameterization. For $1$- and $100$-$\mearth$ planets this is $6.9$ and $5.2$~percent respectively when $\umax=3$. If we had chosen $\umax=1$, the incompleteness of the standard relative to the \croin{} parametrization would have been $21$~percent for both $1$- and $100$-$\mearth$ planets. As we shall see below, the planet detections that the standard parameterization misses are primarily planets with large separations that are likely to be observed as ``free-floating'' planets with no destection of the host's microlensing event. 

\subsection{Biases}\label{biases}

Using the \croin{} parameterization will necessarily bias the results of a simulation, so we must be aware of how each parameter will be affected and the magnitude of any bias. We consider how using the \croin{} parameterization affects the distribution of each of the fundamental binary microlensing event parameters, beginning with $\tzero$

\begin{figure}
\includegraphics[width=\columnwidth]{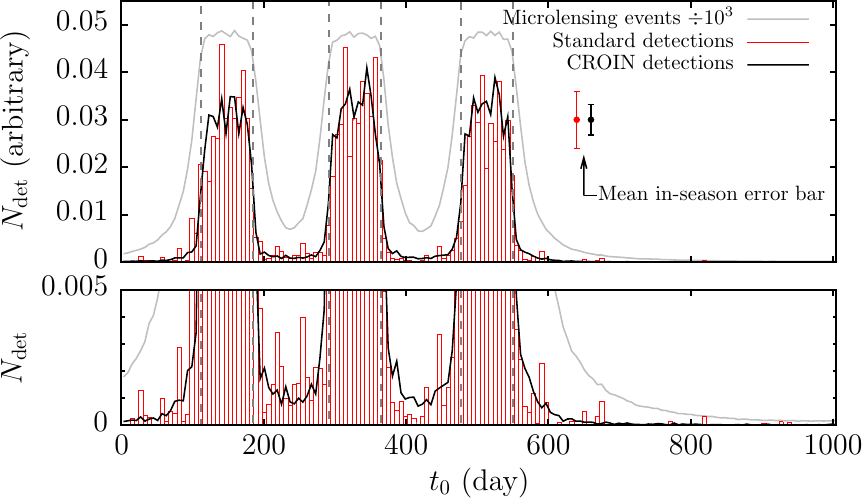}\\
\includegraphics[width=\columnwidth]{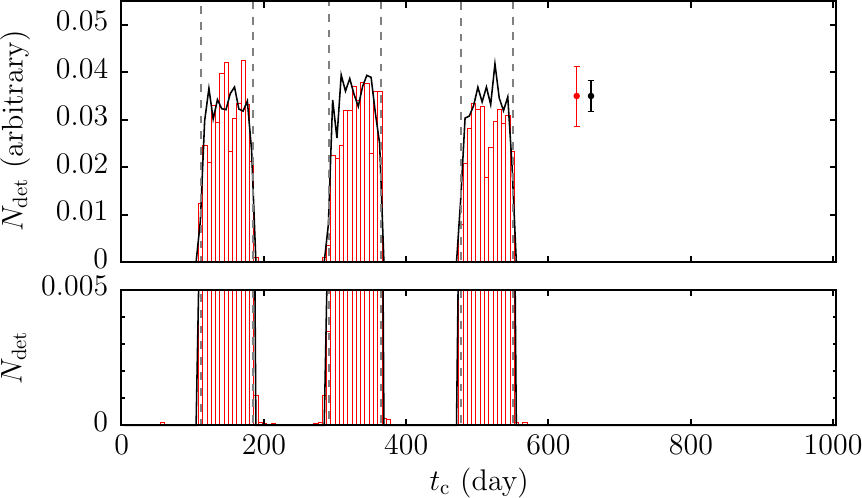}
\caption{\emph{Top:} Red and black lines plot of the number of $1$-$\mearth$ planet detections as a function of the time of closest approach of the source to the primary lens ($\tzero$) for events simulated with the standard and \croin{} parameterizations respectively. The gray line shows the distribution of $\tzero$ for all microlensing events that were detectable in the standard simulation (where detectable means the event caused a $\Delta\chi^2>500$ deviation relative to a flat lightcurve). \emph{Bottom:} The same, but plotted against the time of closest approach to the center of the \croin{} ($\tc$). Lower panels of each plot zoom in on the lowest portion of the main plot to show low-level features. Error bars show the typical uncertainties on the number of detections in the in-season bins (bin-width is 6~d).  Dashed lines show the boundaries of the \wfirst{} seasons, which last $72$~days.}
\label{t0fig}
\end{figure}

Figure~\ref{t0fig} shows the number of planet detections $N_{\mathrm{det}}$ as a function of both $\tzero$ and $\tc$. For the WFIRST survey, 24~percent of detectable microlensing events (where the microlensing event causes a $\Delta\chi^2>500$ deviation relative to a flat baseline) peak outside the observing seasons, and a significant fraction (${\sim}10$~percent) of all planet detections occur in such events. In this plot, the detections from the \croin{} simulation are restricted to the same parameter range as the standard simulation ($|\uzero|<3$, $0<\tzero<2010$~d). The $\tzero$ distributions from each simulation match extremely well, implying that any bias in the $\tzero$ distribution due to the use of the \croin{} parameterization is small. 

The comparison of the $\tzero$ and $\tc$ distributions of planet detections in Figure~\ref{t0fig} demonstrates the value of using the \croin{} parameterization in low-duty cycle simulations. For $1$-$\mearth$ planets just $0.8\pm 0.2$ of planets have $\tc$ falling outside the observing season, whereas if we were to restrict ourselves to only events with $\tzero$ in season to improve the efficiency of our simulations, we would miss $10.5\pm0.7$~percent of planet detections (for $1$-$\mearth$ planets). For $100$-$\mearth$ planets the fractions outside the season are more comparable, due to the longer timescale of the planetary perturbation. In this case $4.8\pm0.8$~percent of detections have $\tc$ fall outside the season, while $9.6\pm1.0$~percent have $\tzero$ fall outside the season.

\begin{figure}
\includegraphics[width=\columnwidth]{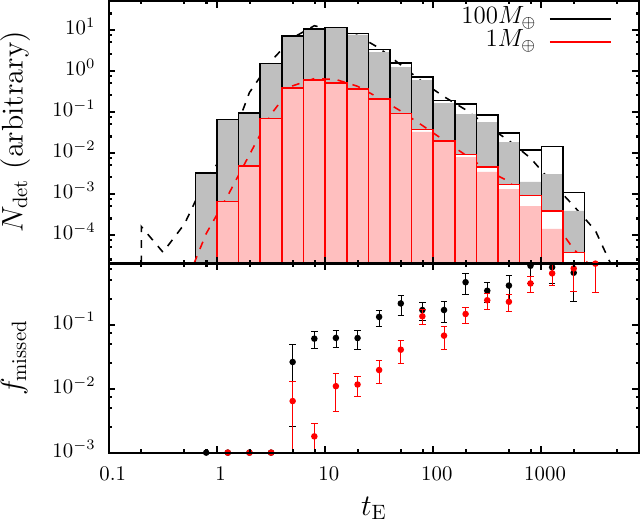}
\caption{\emph{Upper panel:} The event timescale distribution of planet detections for $1$-$\mearth$ (red/pink) and $100$-$\mearth$ (black/gray) planet detections. The histogram drawn with a solid line is the result of the standard simulation. The lighter colored solid fill shows those events in the standard simulation that fall within the \croin{} parameter limits. The dashed line shows the results of the \croin{} simulation. \emph{Lower panel:} The incompleteness $f_{\mathrm{missed}}$ of the \croin{} parameterization relative to the standard simulation. If the incompleteness is less than $10^{-3}$ a point is plotted on the bottom axis (e.g., the points below $\tein=5$~d). Color coding is the same as in the upper panel.}
\label{tecompleteness}
\end{figure}

Figure~\ref{tecompleteness} plots the incompleteness of the \croin{} parameterization relative to the standard parameterization as a function of the event timescale in the bottom panel, and demonstrates the bias introduced by using the \croin{} parameterization in the top panel. Looking at the top panel, we can see that the bias in the timescale distribution introduced by the \croin{} parameterization will be a small change in the slope of the already steeply declining large-timescale tail. However, while the apparent change in shape of the timescale distribution is difficult to notice (by comparing the solid to the dashed line), the incompleteness steadily rises as $\tein$ increases, and approaches $100$~percent for the extremely rare, longest-timescale events.

\begin{figure}
\includegraphics[width=\columnwidth]{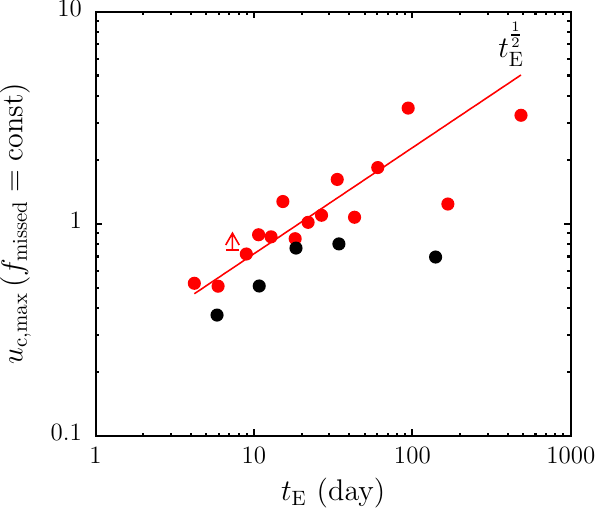}
\caption{Plot of the $\ucmax$ that would be required to maintain the average incompleteness of $1.3$~\% for $1$-$\mearth$ planets (red points) or $6.9$~\% for $100$-$\mearth$ planets (black points) as a function of $\tein$. Arrows represent upper limits. The approximate $\tein^{1/2}$ scaling can be used to adjust $\ucmax$ to account for changes in cadence relative to the $15$~min WFIRST cadence we have simulated, because increasing the timescale will cause the $\Delta\chi^2$ of a planet detection to change in the approximately the same way as increasing the cadence for a fixed mass-ratio planet.}
\label{ucte}
\end{figure}

We can use this trend of incompleteness with the event timescale to judge the impact of changing the cadence of observations, despite running our simulations at a fixed cadence. For an microlensing event with fixed $q$ and fixed source angular diameter relative to the Einstein radius, increasing the timescale by some factor has exactly the same effect on the $\chi^2$ as increasing the cadence by the same factor. We can therefore estimate the necessary change in $\ucmax$ by finding the required $\ucmax$ that would make the incompleteness match that of the whole sample in bins of $\tein$. Figure~\ref{ucte} shows this for both simulations. $\ucmax$ scales as roughly as $\tein^{1/2}$ implying that it should also scale as $f^{1/2}$, where $f$ is the frequency of observations. Of course, we have assumed a fixed mass and not a fixed mass ratio, so this argument may not be strictly valid. However, the mean mass ratio across the entire range of $\tein$ only changes by a factor of ${\sim}2$ compared to the factor of ${\sim}100$ range of $\tein$, so the effect of the changing mean mass ratio is likely to be small. 

\begin{figure*}
\begin{tabularx}{\textwidth}{XX}
\includegraphics[width=0.49\textwidth]{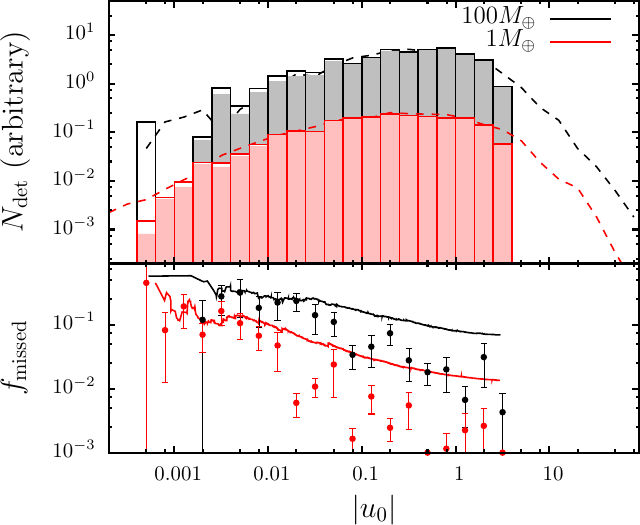}&
\includegraphics[width=0.49\textwidth]{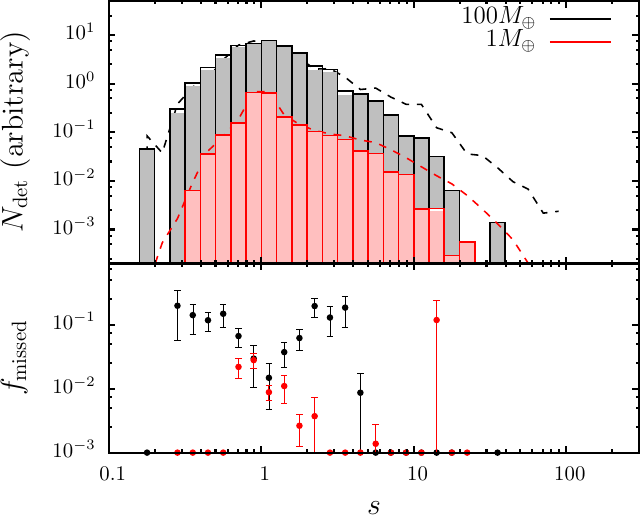}\\
&\\
\includegraphics[width=0.49\textwidth]{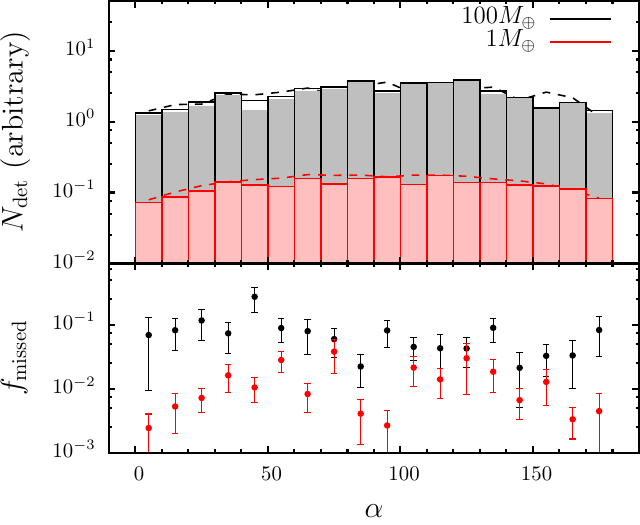}&
\includegraphics[width=0.49\textwidth]{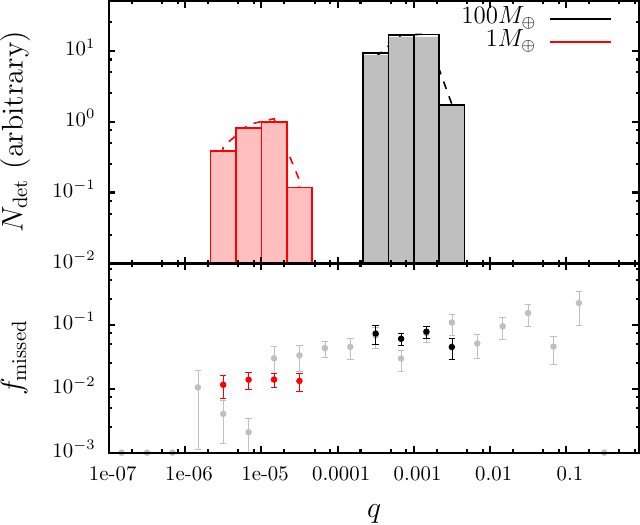}\\
\end{tabularx}
\caption{The same plots as Figure~\ref{tecompleteness} but for the other binary microlensing parameters, $\uzero, s, \alpha$ and $q$ (arranged left to right, top to bottom, respectively). The solid lines in the incompleteness panel of the $\uzero$ plot are the cumulative incompleteness $f_{\mathrm{missed}}(<|\uzero|)$, showing the incompleteness one would suffer if $\umax$ were chosen to be smaller than we have chosen. Gray points in the incompleteness panel of the $q$ plot are computed from the data shown in Figure~\ref{croindef}; note that the semimajor axis range simulated in Figure~\ref{croindef}'s data is larger ($0.1<a<100$~AU, gray points) than that for the test simulations shown with black and red points ($0.3<a<30$~AU).}
\label{paramcompleteness}
\end{figure*}

Figure~\ref{paramcompleteness} shows the distributions of $\uzero, s, \alpha$ and $q$ for the planet detections in the standard and \croin{} simulations. It is the distributions of $\uzero$ and $s$ where the largest differences can be seen between the standard and \croin{} parameterizations. 

Unsurprisingly, the \croin{} parameterization performs poorly for high-magnification events, and the threshold $|\uzero|$ for poor performance gets larger for larger planet masses. For the highest-magnification events, the incompleteness seems to saturate at around $10$~percent for $1$-$\mearth$ planets and ${\sim}20$~percent for $100$-$\mearth$ planets, before falling as a power law as $|\uzero|$ increases with an index that seems to depend on the planet mass. The saturation at $f_{\mathrm{missed}}<1$ for small $\uzero$ is not too surprising, because the \croin{} parameterization deals well with high-magnification events with resonant topologies (when it is centered on the primary lens), and when $s$ is close to $1$ but not resonant the \croin{} can grow large enough to encompass the primary lens. The plot of the cumulative incompleteness shows that for any reasonable choice of $\ucmax$, the total incompleteness of the \croin{} parameterization relative to the standard will not increase significantly, e.g., for $1$-$\mearth$ planets and $\umax=1$, $f_{\mathrm{missed}}=1.57$~percent compared to $1.34$~percent for $\umax=3$.

Use of the \croin{} does not significantly bias the distribution of $s$ compared to using the standard parameterization. However, not using the \croin{} does steepen the tail of the distribution at large $s$. As previously noted however, the majority of these events that the standard parameterization misses will appear to be free-floating planets with no sign of a the host star's microlensing event. The largest incompleteness as a function of $s$ occurs when the goat horns that are seen in Figure~\ref{croindef} caused by high-magnification events cross the $\ucmax$ curve. This occurs at different values of $s$ for different planet masses (with $s$ being further away from $1$ for larger planet masses).

The incompleteness as a function of the source trajectory angle $\alpha$ is not of much interest, but we include it in Figure~\ref{paramcompleteness} for completeness. Note however that for Earth-mass planets, the incompleteness seems to be smallest for trajectories that are either parallel or perpendicular to the binary axis. This pattern is not repeated for $100$-$\mearth$ planets.

The final plot in Figure~\ref{paramcompleteness} shows the distribution of mass ratios for our $1$- and $100$-$\mearth$ simulations, as well as the incompleteness as a function of mass ratio. To extend the range of mass ratios that we consider, and to get a sense of the trend, we also plot the incompleteness as measured from the simulation we used to define $\ucmax$, which is shown in Figure~\ref{croindef}. The range of semimajor axis is smaller in the standard simulation ($0.3<a<30$~AU, red and black points) than the Figure~\ref{croindef} simulation ($0.1<a<100$~AU), but this has a negligible effect due to the strong decline in the number of detections as the separation becomes large or small. The trend is well described by a power law, with incompleteness increasing with $q$. A fit to the data from all the simulations yields the relation
\begin{equation}
f_{\mathrm{missed}}=(1.6 \pm 0.3)\text{ percent} \left(\frac{q}{10^{-5}}\right)^{0.27\pm0.04}.
\label{incq}
\end{equation}

\section{Discussion}\label{discussion}

\subsection{Speed Versus Accuracy}

\begin{table}
\caption{Speed-up for different duty cycles with $0.1<a<100$~AU$^1$.}
\label{speedups}
\begin{tabularx}{\columnwidth}{@{}XcccX}
\hline
\hline
Duty-cycle & 100\% & 80\% & 25\% & Incompleteness$^2$\\
Example & --- & KMTNet & WFIRST & (\%) \\
Mass ($\mearth$) & & & \\
\hline
10000 & 1.4 & 1.7 & 5.5 & 18 \\
1000 & 1.9 & 2.4 & 7.7 & 9.8 \\
100 & 3.7 & 4.6 & 15 & 5.3 \\
10 & 10 & 13 & 41 & 2.8 \\
1 & 31 & 39 & 124 & 1.5 \\
0.1 & 99 & 124 & 397 & 0.8 \\
0.01 & 312 & 390 & 1247 & 0.4 \\
\hline
\end{tabularx}
{\bf Note:} 
$^1$Semimajor axis is distributed logarithmically.\\
$^2$Incompleteness estimated using Equation~\ref{incq} from Section~\ref{biases} assuming an average mass ratio of $8.1\times 10^{-6} (M/\mearth)$.
\end{table}

With estimates of the speed-up from Section~\ref{speedup} and the accuracy from Section~\ref{accbias}, we can now objectively assess the usefulness of the \croin{} parameterization. In Table~\ref{speedups} we summarize both the speed-up and accuracy it is possible to achieve for realistic simulations that will cover a range of semimajor axes and masses. We have computed the speed-up in the same way as in Section~\ref{speedup}, but this time assumed a distribution of semimajor axes in the range $0.1<a<100$~AU and a range of duty cycles representative of realistic microlensing surveys. For $100$-$\mearth$ planets, an incompleteness of ${\sim}5$~percent seems a reasonable incompleteness for most applications, and this can provide a significant speed-up of $4$--$15\times$. A ten percent incompleteness for $1000$-$\mearth$ planets however seems too large, and the speed-up is a factor of $2$ less. The \croin{} parameterization therefore seems most likely to be useful for planets of $100$-$\mearth$ and below. For planets of Earth-mass and below the speed-ups are huge, at the minor cost of an inaccuracy of the order of $1$~percent or less.

The dominant cause of the inaccuracy of \croin{} parameterization is the missing planet detections caused by the central caustic of close and wide separation planets. However, it is relatively easy to use a strategy that does not miss these detections. Rather than use the \croin{} for the origin of the parameterization and drawing $|\uc|<\ucmax$, one would use the standard parameterization, drawing events from its usual limits (e.g., $|\uzero|<3$). Now, one would perform two tests on the parameters: the first would check if was a high-magnification event (with some limit on $|\uzero|$, and the second would convert the standard parameters to the \croin{} parameters and check that the \croin{} parameters were within the limits defined in this paper. If either of the tests were passed, then the lightcurve would be computed, otherwise the event would be assumed to be a non-detection. To do this correctly one would need to perform a similar study to this to determine the $\uzero$ limits for central caustic planet detections. As we were concerned primarily with speeding up low-mass planet simulations, where the number of detections due to non-resonant central caustics is expected to be extremely small we have not conducted this study.

Finally, it should be noted that the speed-up that will actually be achieved may be smaller than that we estimate here if lightcurves that contain detectable planet signatures take longer to generate than ones that do not. This will quite often be the case if one is using the hexadecapole approximation to avoid finite source calculations~\citep{Pejcha2009, Gould2008}. Even so, the actual speed-up is still worthwhile.

\subsection{Potential uses}

The magnitude of speed-up it is possible to achieve with the \croin{} parameterization make its application to microlensing simulations and detection efficiency calculations immediately obvious. The advent of new and proposed microlensing surveys has encouraged a flurry of new work on microlensing simulations~\citep{Shvartzvald2012, Green2012, Penny2013, Spergel2013, Henderson2014, Ipatov2014}. Interpretation of these new high cadence surveys will be significantly easier than for previous observations in survey plus follow-up mode~\citep[e.g.,][]{Gould2010}, but will still require extensive calculations of detection efficiencies over a large parameter space~\citep[e.g.][]{Gaudi2000, Gaudi2002, Tsapras2003, Snodgrass2004}. In fact, the increased cadence and area of the surveys will make such analyses significantly more computationally expensive (though Moore's law will help to a certain extent). A bigger challenge will be presented by the advent of space-based microlensing planet searches. These will provide an order of magnitude more microlensing events to search for planets, but will also measure the lightcurves significantly more accurately, increasing the demands on the lightcurve computations, which are already probably close to maximum efficiency. Maximizing the scientific return of space-based surveys will require extensive simulations to optimize the various aspects that can affect the mission, from hardware to survey design. This has proved challenging so far, with only limited parameter exploration possible for low-mass planets. The \croin{} parameterization represents an important way to broaden the scope of planning for these missions, increasing the size of parameter space that can be explored.

The \croin{} parameterization may also find use in the modeling of individual gravitational microlensing events. As mentioned in Section~\ref{intro} parameterizations centered on the caustics are already in use. The analytic limits on the impact parameters of these parameterizations could be useful in restricting a parameter search. However, as downhill fitting will quickly move a trial solution to a local minimum of the parameter space, any speed-up will be modest. 

\section{Conclusions}\label{conclusion}

We have proposed a parameterization of binary gravitational microlensing applicable to planetary microlensing. We have empirically determined an analytic functional form for the limits of the impact parameter within which the vast majority of planetary detections can be expected. We have shown that by using this parameterization and its analytic limits it is possible to speed-up simulations of planetary microlensing by factors of $10$ to $1000$ depending on the mass of planet being investigated. This comes at a cost of excluding a small percentage of planet detections, though this is smaller than or comparable to the loss due to arbitrary truncation of the space of the standard parametrization.

\section*{Acknowledgments}

I would like to thank Scott Gaudi for helpful discussions, Radek Poleski for pointing out an oversight, Eamonn Kerins, Nick Rattenbury, Shude Mao and Annie Robin whose work has contributed to the development of \mabuls{}, and finally Brett Andrews for his zoomorphic insight. I would also like to thank the anonymous referee for suggestions that improved the clarity of the paper. Some of the computational element of this research was achieved using the High Throughput Computing facility of the Faculty of Engineering and Physical Sciences, The University of Manchester.

\bibliographystyle{mn2e}
\bibliography{librarytmp,apj-jour}

\end{document}